\documentclass[]{spie}  

\usepackage{amsmath,amsfonts,amssymb}
\usepackage{graphicx}
\usepackage[colorlinks=true, allcolors=blue]{hyperref}

\title{Theoretical performance limitations and filter selection based on Fisher information of a computational photonic crystal spectrometer for trace-gas retrieval}

\author[a]{Marijn Siemons}
\author[a,b]{Ralf Kohlhaas}
\affil[a]{SRON Netherlands Institute for Space Research, 2333 CA Leiden, The Netherlands}
\affil[b]{Department of Precision and Microsystems Engineering, Delft University of Technology, 2628 CD Delft, The Netherlands}

\authorinfo{Further author information: (Send correspondence to M.S.)\\M.S.: E-mail: m.siemons@sron.nl}

\begin{document} 
\maketitle

\begin{abstract}
As global climate change severely impacts our world, there is an increasing demand to monitor trace gases with a high spatial resolution and accuracy. At the same time, these instruments need to be compact in order have constellations for short revisit times. Here we present a new spectrometer instrument concept for trace gas detection, where photonic crystals filters replace traditional diffraction based optical elements. In this concept, 2D photonic crystal slabs with unique transmission profiles are bonded on a detector inside a regular telescope. As the instrument flies over the earth, different integrated intensities for each filter are measured for a single ground resolution element with a regular telescope. From this detector data, trace gas concentrations are retrieved. As an initial test case we focused on methane and carbon dioxide retrieval and estimated the performance of such an instrument. We derive the Cram\'er-Rao lower bound for trace-gas retrieval for such a spectrometer using Fisher information and compare this with the achieved performance. We furthermore set up a framework how to select photonic crystal filters based on maximizing the Fisher information carried by the filters and how to use the Cram\'er-Rao lower bound to find good filter sets. The retrieval performance of such an instrument is found to be between 0.4\% to 0.9\% for methane and 0.2\% to 0.5\% for carbon dioxide detection for a 300$\times$300~m$^2$ ground resolution element and realistic instrument parameters.
\end{abstract}


\section{INTRODUCTION}
\label{sec:intro}  
The need for the monitoring of greenhouse gas emission sources motivates the development of earth observation systems with a higher spatial and temporal resolution. For this complete distributed systems with many satellites could be considered, which would require small, low-weight optical imaging spectrometers. Conventional grating spectrometers have intrinsic physical limits on the system size due to the need of large free-space optical path length for the separation of wavelengths. Other solutions, as for example static Fourier-transform spectrometers as developed with NanoCarb \cite{Gousset2019}, suffer less from such limitations but have reduced spectral resolution. Recently, a spectrometer concept was invented \cite{Wang2014,Wang2019} which uses an array of photonic crystal filters with quasi-random spectral transmission functions together with computational inversion (for example compressive sensing) to reconstruct the input spectrum. 

\begin{figure}
    \centering
    \includegraphics[width=1\linewidth]{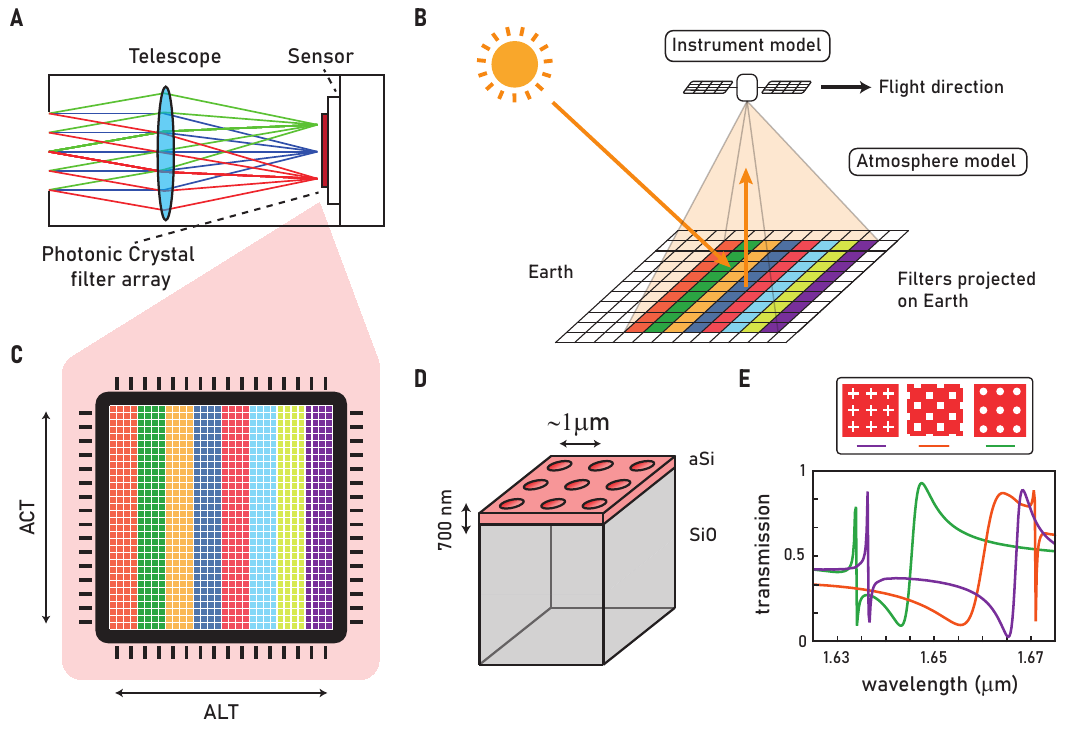}
    \caption{Illustration of the instrument concept. A) The instrument consists of a telescope and sensor. A photonic crystal filter array is placed directly on the sensor. B) As the instrument flies over the earth, ground pixels are observed with different flilters. C) The photonic crystals filter array consists of strips in the across track direction. D) The photonic crystal filters itself is made with a small layer of amorphous silicon (aSi) on top of a quartz substrate. In the aSi-layer 2D patterns are etched, giving rise to distinct and spectrally rich transmission profiles (E). }
    \label{fig:concept}
\end{figure}

Here we introduce a novel instrument concept, using compressive sensing enabled by photonic crystals, see Figure \ref{fig:concept}. The instrument orbits the earth and consists of an earth observing optical telescope, where photonic crystals are positioned on the sensor. These photonic crystals consists of a wafer with a thin dielectric material. In our case this is a glass wafer with a 700 nm thin amorphous Si-film. It contains 2D photonic crystals in a column wise arrangement in the across-track (ACT) direction. These photonic crystals consist of sub-wavelength features etched in the Si-film and have unique spectral transmission profiles. These transmission profiles arise from resonances in the Si-film and depend on the thickness of the film as well as the specific shape and size of the etched structure and the lattice size. By choosing a specific hole shape and lattice size, different spectral transmission profiles can be achieved. Using the push-broom concept, the transmitted intensity for each ground instantaneous field of view pixel can be measured for each spectral filter. From the measured detector values, and the known transmission profiles the trace gas concentration can be estimated using a retrieval algorithm and atmosphere model.

This concept allows a far larger field-of-view in along-track (ALT) direction than with an imaging spectrometer with a slit. This concept has similarities with a linear gradient filter \cite{Blommaert2019}, but the photonic crystal transmission allow for a more diverse transmission profiles and a highly tuned and selective optical system. Moreover, different spectral ranges can be combined on one detector (for example the O2A band and methane band, or several relevant methane spectral ranges). 

A particular interesting advantage is that the photonic crystal filters can be chosen to optimize the information measured by the instrument to tailor trace gas retrieval. This is however very challenging. The photonic crystals can can consist of any different shapes, shape size and lattice spacing, each with their own unique transmission profile. Selecting the best filter set is a non-trivial and difficult optimization problem. Assuming a filter library of around 4000 filters and a maximum of 64 filters to be used, the total amount of combination from the library is 4000 choose 64 = $10^{21}$. This number is so large, all combinations can never be evaluated in linear computational time. Therefore, an algorithm or procedure is needed to find a good performing filter set. 

Fisher information, a concept from mathematical statistics, is a way of quantifying the amount of information a random variable (in this case the measured detector counts) carries about an unknown (fit) parameter. Photonic crystal filters which carry more Fisher information will result in more precise trace gas retrieval. In estimation theory, the Cram\'er-Rao Lower Bound (CRLB) is measure of the best achievable precision for a fit parameter, which can be obtained with an unbiased estimator, i.e. fit algorithm, based on the available Fisher information. In practice, fit algorithms are often biased to some extend and therefore not strictly bounded by the CRLB. 

In this work we derive a lower bound for trace-gas retrieval for this computational photonic crystal spectrometer using the Fisher information and compare this with the achieved performance. We furthermore set up a frame work how to select photonic crystal filters based on maximizing the Fisher information carried by the filters and how to use the CRLB to find good filter sets. A filter set with a low CRLB is however to guarantee that the retrieval algorithm will achieve this precision. Therefore, the CRLB can only be used to find good candidate filter sets, but actual achieved trace gas retrieval precision is used for final selection of a filter set.

The outline of this proceeding is as follows. First we describe the instrument and atmosphere model and the trace gas retrieval algorithm. Next, the photonic crystal simulations are discussed in section 3. Then the Fisher information is derived for the instrument and the filter selection procedure is described in section 4. In section 5 we present the results and selected filter set and compare the achieved retrieval error to the CRLB. Furthermore, the performance of the instrument in different signal to noise scenario's is presented. Lastly, we discuss the implications of this selection method and suggest improvements.

\section{INSTRUMENT AND ATMOSPHERE MODEL}
\subsection{Instrument model}
As described before, the instrument consists of a telescope design and a photonic crystal array positioned close to the detector. As the instrument flies over the earth, each on ground pixel is observed via a number of photonic crystal filters.
The expected intensity $\mu_k$ for filter $k$ is given by
\begin{equation}
    \mu_k =  t_\mathrm{int} G N_\mathrm{pxl}\int_{\lambda_\mathrm{min}}^{\lambda_\mathrm{max}} QE(\lambda) T_k(\lambda) S(\lambda) d\lambda
\end{equation}
with $t_\mathrm{int}$ the integration time per image, $G$ the etendue of the instrument, $QE$ the quantum efficiency of the detector, $T_k$ the filter transmission function, $N_\mathrm{pxl}$ the number of pixels with $k$ and $S$ the spectral radiance of the earth.

Assuming nadir observation, the etendue $G$ is computed as 
\begin{equation}
    G = \frac{A \times \mathrm{giFOV}^2}{h^2}
\end{equation}
with $A$ the aperture size in m$^2$, giFOV the  ground-instantaneous field of view and  of the earth and $h$ the altitude. Here an altitude of 500km (low earth orbit, LEO) is assumed. 
As a detector the Lynred Snake InGaAs detector is used in low-gain and integrate-while-read mode. This detector has 512$\times$640 pixels, a full well capacity of $1.4\cdot10^6 e^-$, a pixel size of 15 $\mu$m and we  assume a QE of 85\%. The focal length (125~mm) is chosen such that the giFOV is 60$\times$60 m$^2$ and the integration time (34~ms) is set such that the smearing will end up with an along-track pixel size of 300~m. 5 across track pixels are binned to end up with a 300$\times$300~m$^2$ resolution element and a total across track swath of 30~km. This design configuration results in a F-number of 6.1 and an etendue of $5.8\cdot10^{-12}$~m$^2$sr.

\subsection{Atmosphere model}
The earth radiance spectrum is simulated using a non-scattering radiative transfer model for the atmosphere, assuming Lambertian reflectance at the surface of the earth. The total column densities are modelled by scaling a standard atmosphere profile \cite{united1976,Gloudemans2008}. Cloud properties are not retrieved and atmospheric scattering is ignored. The earth spectral radiance $S$ as function of wavelength $\lambda$ is given by
\begin{equation}
    S(\lambda) = \frac{ \cos(\zeta_s) }{\pi}E(\lambda) A(\lambda) \exp \left[-r_\mathrm{air}\tau_\mathrm{vert} (\lambda) \right]
\end{equation}
with $\zeta_s$ the sun zenith angle, $E$ the sun's irradiance, $A$ the albedo of the earth, $r_\mathrm{air}$ the air mass factor \cite{Young:94} and $\tau_\mathrm{vert}$ the vertical optical thickness. In these simulations we assume a sun zenit angle of 45\textdegree{}.
Using discretization, the vertical optical thickness is computed as
\begin{equation}
    \tau_\mathrm{vert} = \sum_m \chi_m \sum_l  \sigma_m (\lambda, z_l) c_m (z_l)
\end{equation}
with $\chi_m$ an scaling/enhancement factor in the overall concentration of the molecule $m$. This scaling factor maintains the (vertical) column number density profile $c_m(z_l)$ and $\sigma_m$ the absorption cross-section. The absorption cross-section is computed from the line intensity (HITRAN database \cite{Rothman2021}) and the Voigt line shape corresponding to the temperature and pressure of each layer on a fine wavelength sampling \cite{Gloudemans2008}.

The earth's albedo is described as a polynomial with degree $a$
\begin{equation}
    A(\lambda) = \sum_a A_a \tilde{\lambda}^a
\end{equation}
where $\tilde{\lambda}$ is the normalized wavelength
\begin{equation}
    \tilde{\lambda} = \frac{2\lambda - \left(\lambda_\mathrm{min} + \lambda_\mathrm{max}\right)}{\lambda_\mathrm{max} - \lambda_\mathrm{min}}.
\end{equation}

\begin{figure}
    \centering
    \includegraphics[width=0.75\linewidth]{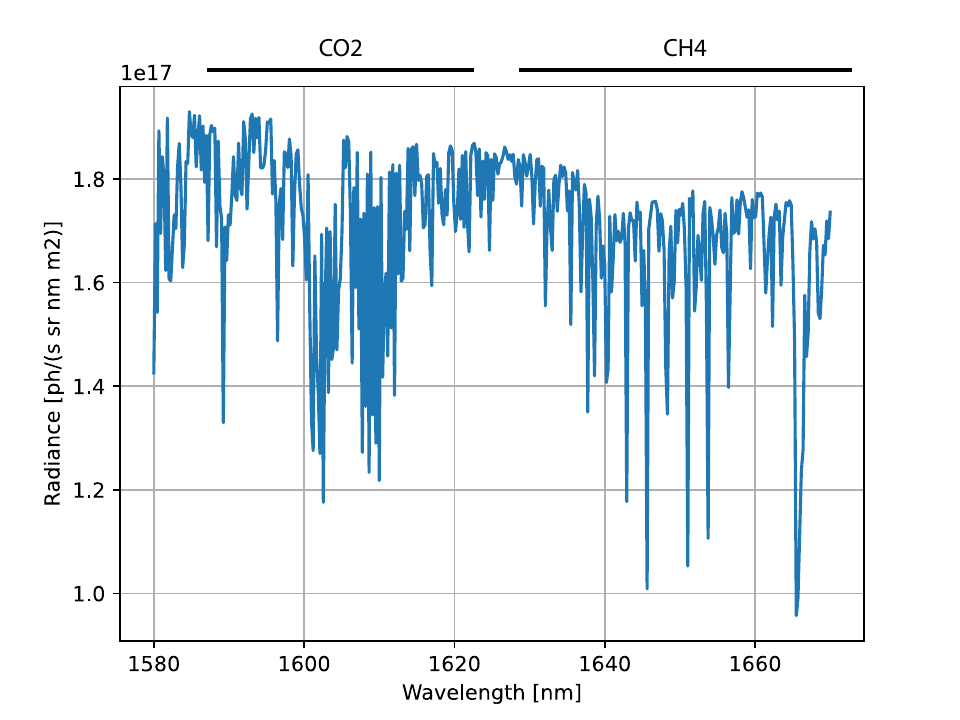}
    \caption{Modeled radiance as function of wavelength. Bars on top indicate the most prominent absorption features of CO2 and CH4.}
    \label{fig:radiance}
\end{figure}

The radiance spectrum associated with the standard profile is shown in Figure \ref{fig:radiance}. CO2 has absorption features between 1590--1620~nm and CH4 has features at 1630--1670~nm and in particular around 1665~nm. 

\subsection{Trace gas retrieval and performance}
In this work we investigate the performance of CH4 and CO2 retrieval in combination with a linearized albedo. This results in fit parameters $\theta$ for the enhancement factors of CH4 and CO2 and and the linearized earth's albedo
\begin{equation}
    \theta = \begin{bmatrix}
\chi_\mathrm{CH4}\\ 
\chi_\mathrm{CO2}\\ 
A_0\\
A_1
\end{bmatrix}
\end{equation}

The trace gas concentrations and albedo are estimated using a least-squares fit
\begin{equation}
    \min_\theta \left(n_k -\mu_k({\theta})\right)^2
\end{equation}
with $n_k$ the measured photon count for filter $k$ and $\mu_k$ the expected model photon count and $\theta$ the fit parameter vector. The least-square fit is efficiently performed with the function \emph{optimize.least\_squares} and `dogbox' method from the python \emph{scipy} library.

For the performance of the instrument, both the achieved precision as well as bias (accuracy) of the retrieval are important. A metric which combines both the bias and the precision is the Root Mean Square error (RMSE) of fit parameter $i$
\begin{equation}
    \mathrm{RMSE}_i = \sqrt{\frac{1}{N}\sum_n \left(\theta^n_i-\theta_i\right)^2}
\end{equation}
with $\theta^n_i$ the fitted value of noisy realization $n$, $\theta_i$ the ground truth value and $N$ the total amount of noisy realization. Alternatively, the RMS error can be seen as a orthogonal combination of the bias and standard deviation (precision)
\begin{equation}
    \mathrm{RMSE}_i = \sqrt{\sigma_i^2 + b_i^2}
\end{equation}
with $b_i = \bar{\theta}_i-\theta_i$ the bias, $\sigma_i = 1/(N-1)\sum_n(\theta^n_i-\bar{\theta}_i)^2$ the standard deviation and $\bar{\theta}_i$ the sample mean.

\section{Photonic crystal simulations }
\begin{figure}[b]
    \centering
    \includegraphics[width=0.5\linewidth]{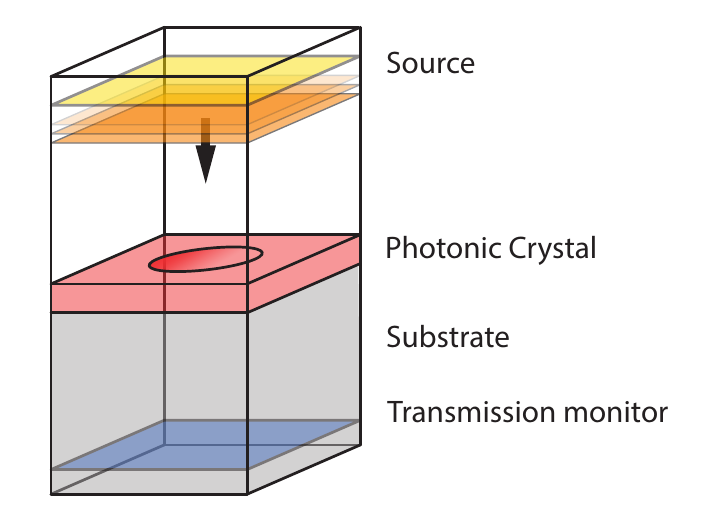}
    \caption{Illustration of the photonic crystal simulation. It consists of a aSi-layer on top of a quartz substrate. }
    \label{fig:simulation-illustration}
\end{figure}

\begin{figure}
    \centering
    \includegraphics[width=\linewidth]{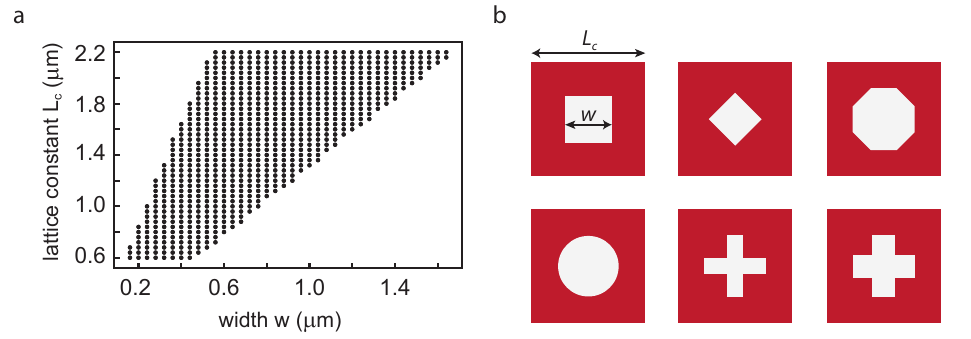}
    \caption{Photonic crystal library consists of difference sized structure, both lattice constant and size of the structure (a) for 6 different shapes (b).}
    \label{fig:simulation-library}
\end{figure}

The photonic crystals are simulated with Lumerical (ANSYS) using Rigorous Coupled Wave Analysis (RCWA), see Figure \ref{fig:simulation-illustration}. The simulation domain consists of quartz substrate with a 700~nm layer of amorphous silicon (aSi) on top. For simulation parameters a wavelength spacing of 0.2~nm is chosen over a range of 1580--1670~nm and the amount of spatial vectors (the number of spatial modes per layer) is set to 100, which seems to be a good compromise between accuracy and computational speed. We computed the transmission profile for 6 different shapes with a periodicity ranging from 600~nm to 2.2~\textmu m with 40~nm steps, see Figure \ref{fig:simulation-library}. These RCWA simulation in Lumerical provide directly the transmission information and a total library of 4500 filter was simulated.

\section{Fisher Information and CRLB derivation}
\subsection{Fisher information in the presence of shot and read noise}
The derivation of the general Fisher information matrix and the associated log-likelihood function, which includes shot noise and read noise has been described before \cite{Chao2012,Huang2013}. Here a comprehensive derivation is provided. 
Let $\mu_k$ be the expected  photoelectron count ($e^-$)  and $n_k$ the observed photoelectron count for filter $k$. The probability density function $P_k$ which includes both shot noise and read noise can be described as a discrete convolution between a Poisson distribution (shot noise) and a Gaussian distribution (read noise) \cite{Huang2013}
\begin{equation}
    P_k(\mu_k | n_k) = A \sum_{q=0}^{\infty} \frac{1}{q!}e^{-\mu_k}n_k^q \frac{1}{\sqrt{2\pi \sigma_k^2}} \exp\left[\frac{(\mu_k-q)^2}{2\sigma_k^2}\right]
    \label{fullPDF}
\end{equation}
with $A$ the normalization constant, $q$ the convolution parameter and $\sigma_k = N_\mathrm{pxl}\sigma_\mathrm{pxl}$ where $N_\mathrm{pxl}$ is the total number of pixels with the associated filter and $\sigma_\mathrm{pxl}$ the RMS read noise from a single pixel.
Since a Gaussian distribution asymptotically approaches a Poisson distribution and since the convolution of two Poisson distribution with mean $a$ and $b$ is equal to a Poisson distribution with mean $a+b$, equation \ref{fullPDF} can be approximated by
\begin{equation}
    P_k (\mu_k | n_k )= \frac{\left(\mu_k + \sigma_k^2\right)^{n_k+\sigma_k^2}}{\Gamma\left(n_k+\sigma_k^2+1\right)} \exp\left[-\left(\mu_k+\sigma_k^2\right)\right]
\end{equation}
with the gamma-function $\Gamma(n+1) = n!$. The log-likelihood function $\log L$, which is the natural log of the probability density function, is given by
\begin{equation}
    \log L  = \sum_k \left[ \left(n_k + \sigma_k^2\right) \log \left(\mu_k + \sigma_k^2\right) - \left(\mu_k + \sigma_k^2\right) - \log \Gamma \left(n_k + \sigma_k^2 + 1\right) \right].
\end{equation}
The derivatives of the log-likelihood w.r.t. the fit parameters $\theta$ are given by
\begin{equation}
    \frac{\partial \log L}{\partial \theta_i} = \sum_k\frac{n_k-\mu_k}{\mu_k + \sigma_k^2} \frac{\partial \mu_k}{\partial \theta_i}
\end{equation}
and the Fisher information matrix is found to be
\begin{equation}
    F_{jl} = \left \langle \frac{\partial \log L}{\partial \theta_j} \frac{\partial \log L}{\partial \theta_l}  \right \rangle = \sum_k \frac{1}{\mu_k + \sigma_k^2} \frac{\partial \mu_k}{\partial \theta_j} \frac{\partial \mu_k}{\partial \theta_l}
    \label{eq:fisher-1}
\end{equation}
assuming that the photon counts of each filter $k$ are uncorrelated to each other. This Fisher matrix is a general solution for a system with shot and read noise. 

To compute the Fisher matrix of the photonic crystal spectrometer the derivative of $\mu_k$ with respect to fit parameter $\theta_j$ is needed, which given by
\begin{equation}
    \frac{\partial \mu_k}{\partial \theta_j} = t_\mathrm{int} G \frac{N_\mathrm{pxl}}{N_\mathrm{filter}} \int_{\lambda_\mathrm{min}}^{\lambda_\mathrm{max}} QE(\lambda) T_k(\lambda) \frac{\partial S(\lambda)}{\partial \theta_{j}} d\lambda.
\end{equation}
For trace gas enhancement factor $\chi_m$ of molecule $m$, the partial derivative ${\partial S(\lambda)}/{\partial \theta_{\chi_m}}$ is given by
\begin{equation}
    \frac{\partial S(\lambda)}{\partial \theta_{\chi_m}} =  -S(\lambda) r_\mathrm{air} \sum_l  \sigma_m (\lambda, z_l) c_m (z_l)
\end{equation}
and the derivative of $S(\lambda)$ with respect to the albedo coefficient $A_a$ is given by
\begin{equation}
    \frac{\partial S(\lambda)}{\partial \theta_{A_a}} =  \frac{ \cos(\zeta_s) }{\pi}E(\lambda) \exp \left[-r_\mathrm{air}\tau_\mathrm{vert} (\lambda) \right] \tilde{\lambda}^a = \frac{S(\lambda)}{A(\lambda)}  \left(\frac{2\lambda - \left(\lambda_\mathrm{min} + \lambda_\mathrm{max}\right)}{\lambda_\mathrm{max} - \lambda_\mathrm{min}}\right)^a. 
\end{equation}
The Cram\'er-Rao Lower Bound of each fit parameter is the appropriate diagonal element of the inverse Fisher matrix
\begin{equation}
    \mathrm{CRLB}_j = \sqrt{{F^{-1}}_{jj}}.
\end{equation}

The filters used in the instrument are represented by the selection $k$ and the Fisher information matrix is the sum of the individual contribution of each filter. Alternatively to eq. \ref{eq:fisher-1} , the Fisher information can be described as the sum of the individual contribution of each filter
\begin{equation}
    F_{jl} = \sum_k \frac{1}{\mu_k + \sigma_k^2} \frac{\partial \mu_k}{\partial \theta_j} \frac{\partial \mu_k}{\partial \theta_l} = \sum_k s_j^k s_l^k
\end{equation}
with $s_k$ the score vector of filter $k$, given by
\begin{equation}
    s^k_i = \frac{1}{\sqrt{\mu_k+\sigma^2_k}} \frac{\partial \mu_k}{\partial \theta_i} .
\end{equation}
This vector describes the amount of Fisher information each filter $k$ is contributing. The score vector can be interpreted as the ratio between the sensitivity of the filter $k$ to the fit parameter $\theta_i$ and the noise in the measured value for filter $k$. 

\section{Filter selection by Fisher information optimization}
For optimal performance of the photonic crystal spectrometer, a good filter set has to be found which achieves the best retrieval performance. To reduce the electromagnetic interactions at the boundaries of the filters, we expect that as maximum 64 filter can be placed on the detector (120~\textmu{}m per filter). 16 filters are assumed to be the minimum to ensure redundancy in the retrieval algorithm. As mentioned before, finding a good or the best filter set is a computational challenging optimization problem as the number of possible combinations exceeds $10^{21}$. Assuming an evaluation time of 1~ms per set, the total duration would be 30 billion years. Therefore, an alternative method is needed.

Fisher information can be used to leverage this problem. The Fisher information matrix entries for each filter $F_{ij}^k$ can be pre-computed since these don't rely on the specific filter combination. Therefore, the Fisher matrix and CRLB can be calculated for different filter sets by a simple sum and straight forward matrix inversion for only the diagonal elements, which together are only a handful floating point operations.

\begin{figure}
    \centering
    \includegraphics[width=0.5\linewidth]{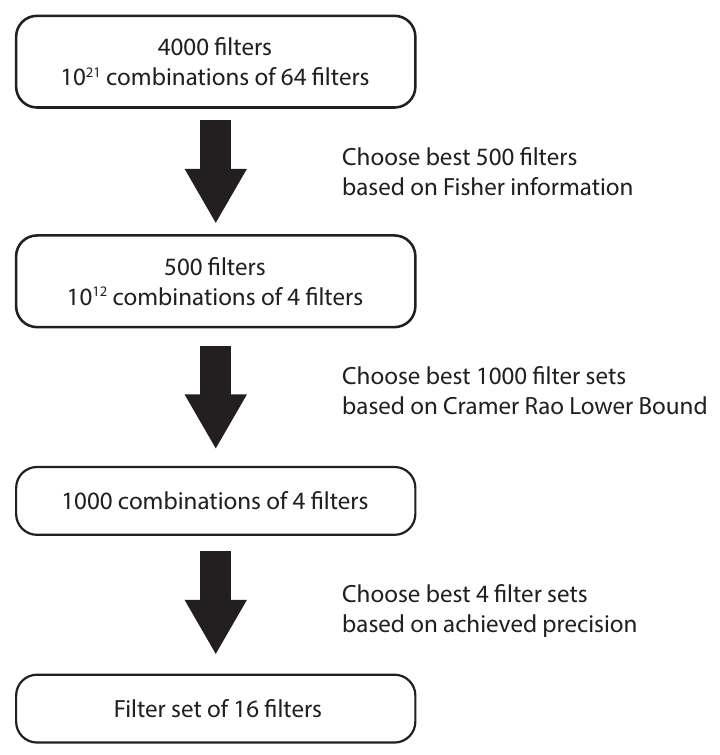}
    \caption{Diagram of the filter selection procedure.}
    \label{fig:filter_selection_procedure}
\end{figure}

We propose the following procedure filter selection procedure, see Figure \ref{fig:filter_selection_procedure}. In short, first a pre-selection is made using the Fisher Information as ranking. Then the CRLB is evaluated for all filter set combinations and the best 1000 filter sets are selected. Lastly, the achieved performance (RMS error) of these filter sets based on simulated noisy retrievals is assessed and the best filter sets are chosen and combined. This results in a final combined filter set of 16 filters. We now describe the method in more detail.
\subsection{Filter pre-selection}
First, filters are pre-selected from the library. Many filters do not exhibit interesting features and can be excluded in the assessment. The filters are ranked by the relative score vector, given by
\begin{equation}
    \widetilde{s}^k_i = \frac{s^k_i}{\frac{1}{K}\sum_k s^k_i}.
\end{equation}
This is a measure of how much information filter $k$ carries relative to average of the library. This normalization ensures that the weight of each fit parameter is equal. Filters which have a short score vector length provide little information for the trace gas retrieval, relative to the average and filters with a large score provide more information. The length of the relative score vector is therefore an excellent measure to pre-select filters. Here we selected 500 filters with the largest Fisher vector length. 

\subsection{CRLB evaluation}
Next the CRLB is evaluated for all 4 filter combinations which can be made from the 500 filters. This results in a total of 500 choose 4= 2.5 billion filter sets consisting of 4 different filters. Expanding the number of filters in the combination, to for instance 8 or 16, would lead to too many options to assess. Therefore the number of combination is chosen to be 4 and then later combine the sets. To efficiently evaluate all filter set combination the entries of the Fisher Information matrix are pre-computed. Next all the CRLB values for methane and carbon for all the 2.5 billion filter set combinations are computed on the GPU with a custom GPU kernel. This kernel is able to compute all CRLB evaluations on a laptop GPU (NVIDIA T500) under 10 seconds.

\subsection{Performance assessment}
Lastly, the performance of the 1000 best filter sets based on the CRLB are assessed and the best 4 filter sets, which each consisted of 4 filters, are chosen. This results in a filter set of a total of 16 filters. Multiple filter sets are combined to ensure redundancy in the system and increase the robustness of the retrieval. 

\section{Simulation results}
\begin{figure}
    \centering
    \includegraphics[width=0.75\linewidth]{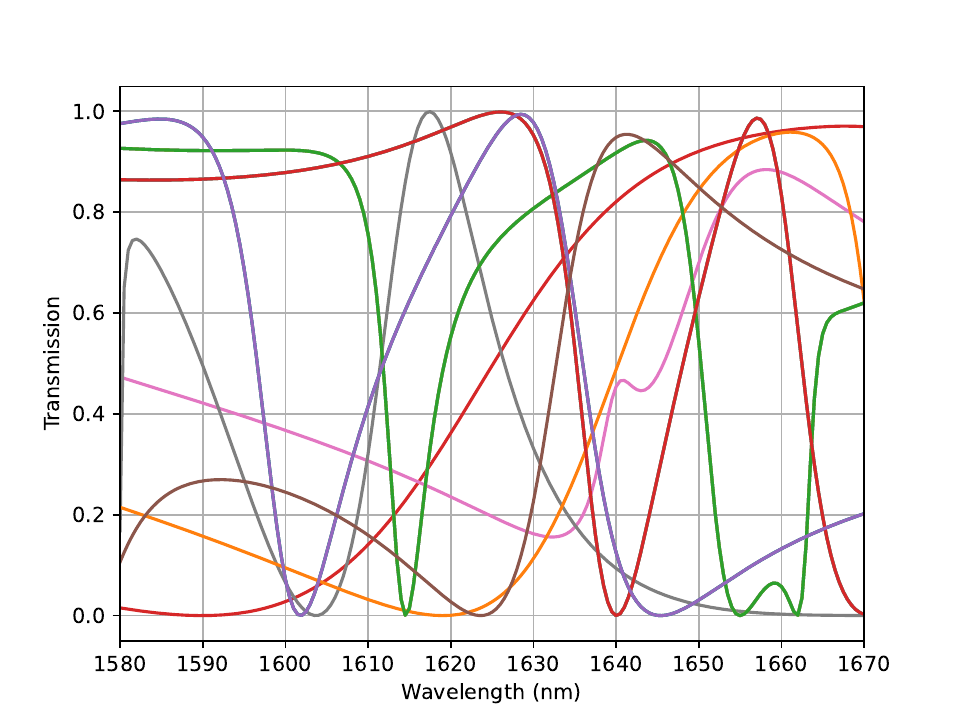}
    \caption{Transmission profiles of the selected filter set.}
    \label{fig:selected_filters}
\end{figure}

The selected filters by the filter selection procedure are shown in Figure \ref{fig:selected_filters}. The transmission profiles show spectral features with high contrast, some with multiple narrow band features and some smoother high contrast features covering the complete band. Clearly these features result in a favourable CRLB and therefore performance.
The filter set presented in Figure \ref{fig:selected_filters} consists of 4 filter sets, each containing 4 filter. These 4 filter sets do have some overlap in selected filters: filters with ID 151 and 2380 are present 4 times and filter 4023 is present 3 times. If needed, the filter selection method can be modified to force only unique solution, but some degree of overlap is likely acceptable. 

Surprisingly, when comparing the CRLB of the individual filter sets to the combined filter set, the CRLB actually decreases. Therefore the combination of the different filter sets improves the performance even further. A filter selection procedure which is able to assess these larger sets directly would likely improve the performance further as it might be able to find the better combinations. Unfortunately the number of different combinations with 8 or 16 filter again becomes so large that it cannot be evaluated anymore using this procedure. Therefore for these larger filter sets a different approach is needed.

\begin{figure}[h!]
    \centering
    \includegraphics[width=\linewidth]{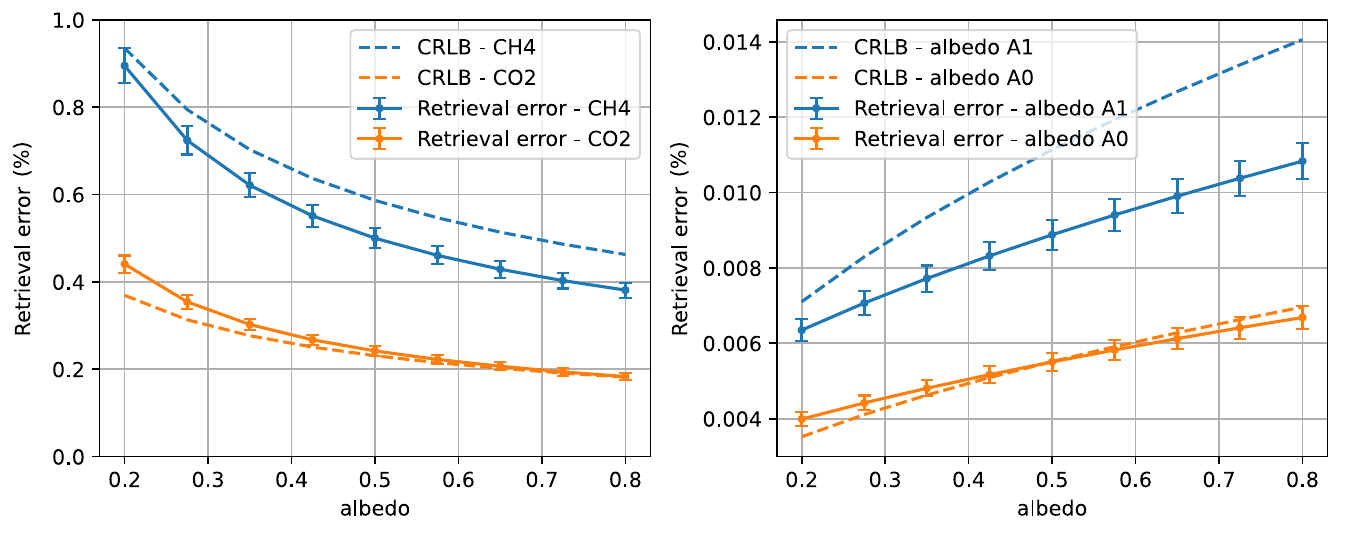}
    \caption{Achieved retrieval error as function of albedo $A_0$ for a concentration of 1800 ppb for CH4 and 330 ppm for CO2. Dashed line indicates the CRLB. Left: Retrieval error of CH4 and CO2. Errorbars indicate the standard deviation for N=250 simulations. Right: Retrieval error of albedo coefficients.}
    \label{fig:retrieval_error}
\end{figure}

\begin{figure}[h!]
    \centering
    \includegraphics[width=\linewidth]{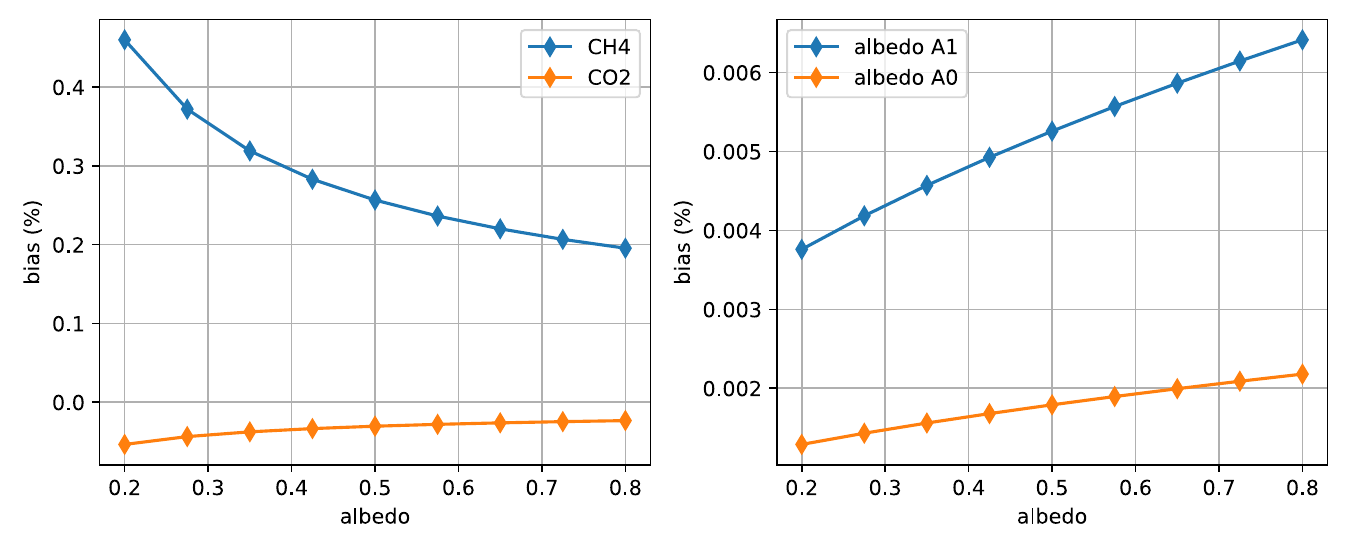}
    \caption{Left: Relative retrieval bias of CH4 and CO2. Right: Relative retrieval bias of albedo coefficients.}
    \label{fig:retrieval_bias}
\end{figure}

Next we compared the CRLB to the achieved retrieval performance, for both the trace gas retrieval as the albedo retrieval, shown in Figure \ref{fig:retrieval_error}. These simulations reveal an retrieval error between 0.5\% to 1\% for methane and 0.2\% to 0.5\% for carbon dioxide detection depending on the albedo coefficient, for realistic instrument parameters. The retrieval error appears to match the CRLB very well for CO2 and for CH4 the performance appears to be slightly lower than the CRLB. The albedo retrieval matches well CRLB for the constant term, but the linear term again outperforms the CRLB. Figure \ref{fig:retrieval_bias} shows the retrieval bias and reveals that for both cases where the instrument outperforms the CRLB a bias is present. This explains the discrepancy observed between the CRLB and the performance.

\section{CONCLUSION}
In this work we show the Fisher information and CRLB derivation and theoretical performance limits for an instrument concept with photonic crystal filters in the focal plane. The retrieval performance of such an instrument is found to be between 0.4\% to 0.9\% for methane and 0.2\% to 0.5\% for carbon dioxide detection for a 300$\times$300~m$^2$ ground resolution element and realistic instrument parameters. The proposed filter selection method is able to find good solutions in a brute-force manor by reducing the optimization space and an efficient computational implementation. The selection method is likely to find the best filter set, but is limited to 4 filters in a single filter set. Increasing the filter set to 8 instead of 4 increases the amount of options by a factor $10^7$, which cannot be brute-force evaluated any more. An alternative method to select filter sets which would allow for larger filter sets (16 or 32 instead of 4) is using an evolutionary algorithm (EA) and the achieved precision as optimization metric. The EA can use the difference in Fisher score of each filter as a mutation rate to efficiently probe the optimization space with filters which provide similar Fisher information. We expect that an additional 10--20\% performance improvement can be achieved in this way for the current library.
The retrieval algorithm is able to retrieve methane concentrations at the CRLB for CO2 and outperforms the CRLB in the case of CH4. This is contributed to the small bias which is observed in the retrieval. Therefore the derived CRLB appears to be a excellent performance indicator.

The results presented here are based on a plane-wave transmission of the photonic crysals, while in practice light is focused onto the photonic crystal slab and detector. The focused beam transmission is expected to be deviate from the plane-wave transmission, but there is currently no established method for performing the instrument design trade-off regarding F-number, pupil shape and aberration sensitivity. The presented filter selection method is however fast enough ($<$1h) to perform on many different filter libraries for different F-numbers and pupil shapes, if these are computed. Therefore, we believe this filter selection method will a be highly useful method for the instrument design phase. 

\acknowledgments 
The authors would like to thank Johannes Algera, Jochen Landgraf and Paul Tol for help on the atmosphere model and constructive discussion and feedback. 

\bibliography{report} 
\bibliographystyle{spiebib} 

\end{document}